\documentclass[journal,letterpaper,10pt,twocolumn,twoside]{IEEEtran}

\IEEEoverridecommandlockouts

\usepackage{times,amsmath,epsfig,amssymb,graphicx,amsfonts,amsthm}
\usepackage{empheq}
\usepackage{psfrag}
\usepackage{fge}
 \usepackage{dsfont}
 \usepackage{dblfloatfix}
 \usepackage{tikz}
\usepackage{amsmath}
\usepackage{amsthm}
\usepackage{amsfonts}   
\usepackage{amssymb}
\usepackage{mathrsfs}
\usepackage{setspace}
\usepackage{caption}
\usepackage{subcaption}
\usepackage{algorithm}
\usepackage{algcompatible}
\usepackage{enumerate}
\usepackage{stmaryrd}
\usepackage{mathrsfs}
\usepackage{mathtools}
\mathtoolsset{showonlyrefs}

\usepackage{soul}

\floatname{algorithm}{Routine}

\newtheorem{mydef}{Definition}
\newtheorem{mytheorem}{Theorem}
\newtheorem{claim}{Claim}
\newtheorem{mylemma}{Lemma}
\newtheorem{myremark}{Remark}
\newtheorem{mycorollary}{Corollary}

\newtheorem{myexample}{Example}
\newtheorem{myproblem}{Problem}

\newcounter{ale}

\newenvironment{liste}{\begin{itemize}}{\end{itemize}}
\newcommand{\aliste}{\begin{liste} \setcounter{ale}{1}}
\newcommand{\zliste}{\end{liste}}

\title{ Detection and Isolation of Failures in Directed Networks of {LTI} Systems}

\author{Mohammad Amin Rahimian, Victor M. Preciado{\small $~^{*}$}
\thanks{$^{*}$ The authors are with the Department of Electrical and Systems Engineering, University of Pennsylvania, Philadelphia, PA 19104-6228 USA. (email: {\fontsize{8}{8}\selectfont\ttfamily\upshape preciado@seas.upenn.edu}). 

This work was supported by the National Science Foundation grants CNS-1302222 ``NeTS: Medium: Collaborative Research: Optimal Communication for Faster Sensor Network Coordination'', and IIS-1447470 ``BIGDATA: Spectral Analysis and Control of Evolving Large Scale Networks''.}}

\begin{document}
\maketitle

\begin{abstract}

We propose a methodology to detect and isolate link failures in a weighted and directed network of identical multi-input multi-output LTI systems when only the output responses of a subset of nodes are available. Our method is based on the observation of jump discontinuities in the output derivatives, which can be explicitly related to the occurrence of link failures. The order of the derivative at which the jump is observed is given by $r(d+1)$, where $r$ is the relative degree of each system's transfer matrix, and $d$ denotes the distance from the location of the failure to the observation point. We then propose detection and isolation strategies based on this relation. Furthermore, we propose an efficient algorithm for sensor placement to detect and isolate any possible link failure using a small number of sensors. Available results from the theory of sub-modular set functions provide us with performance guarantees that bound the size of the chosen sensor set within a logarithmic factor of the smallest feasible set of sensors. These results are illustrated through elaborative examples and supplemented by computer experiments.

\end{abstract}

\section{Introduction}

Fault Detection and Isolation (FDI) is an active area of research
with a wide range of applications, such as power systems analysis
\cite{largeScaleNonlinearPowerNetworks}, robotic networks \cite{mesbahiBook},
and security of cyber-physical systems \cite{6545301}. {In
particular, reliability analysis of} multi-agent networks {is
vital in many areas of engineering, since many critical infrastructures
can be modeled as networked dynamical systems}. {In
this context, the collective} dynamics of a network of dynamic agents
can be severely affected by network failures, resulting
in undesirable behaviors. Hence, studying the effects of link and/or
node failures on the network dynamics is of vital importance with
a wide range of practical implications \cite{aminAutomatica,Kleinberg}. On account of its relevance, there is a wide literature on (Failure Detection and Isolation) FDI techniques and engineering applications (see, for example, \cite{chen2012robust} and references therein).

{The analysis and development of FDI techniques for
networks of LTI systems is theoretically, as well as practically,
appealing. Networks of LTI systems have been extensively investigated,
since they encompass particularly relevant dynamics, such as linear agreement protocols \cite{ConsensusOlfatiTAC2004,RenBeard2008}. In this paper, we focus our attention to this class of systems and develop graph-theoretic tools for FDI.} Our work is related to \cite{Yoon:2011:TFR:2286661.2286741}, in which Yoon and Tsumura
use graph theory to draw interesting results about the stability margins
of transfer functions between pairs of input/output nodes in terms
of the length of shortest paths. In \cite{asjc}, Rahimian
et al. derive sufficient conditions for detectability and
identifiability of link failures in
terms of inter-nodal distances between link failures and observation
points.

The main objective of this paper is to provide an explicit methodology
for FDI in directed networks of LTI agents. Our
algorithms are based on the analysis of discontinuities in the derivatives
of the output responses of a subset of sensor nodes. Based on our
results, we also provide an efficient sensor placement algorithm {for FDI using a small number of sensors.
Although finding the minimum number of sensors for FDI is a hard combinatorial
problem, we provide an approximation algorithm with quality guarantees
based on submodular set functions. In particular, we prove that the our
algorithm is a $\log\left|\mathcal{E}\right|+1$ approximation to
the combinatorial FDI problem, where $\left|\mathcal{E}\right|$ is the number of edges in the network.}

The remainder of this paper is organized as follows. Section~\ref{sec:pre} begins with some preliminaries
on graph and matrix theory, as well as the networked dynamics. In Section~\ref{sec:formulation}, we introduce the model
of dynamic network under consideration and formulate the detection
and isolation problem. Section~\ref{sec:main} contains the relationship between the location
of link failures and discontinuities in the network output signals.
In Subsections \ref{sec:coverage} and \ref{sec:resolution}, efficient
algorithms are proposed to find an effective selection of observation
nodes, for both detection and isolation problems. In Subsection \ref{sec:setcover},
we use results from the theory of submodular set functions to derive performance guarantees for our algorithms. Illustrative examples and discussions in Section~\ref{sec:examples} elucidate the results followed by computer experiments on large random networks. Section~\ref{sec:conc} concludes the paper. All proofs are included in the Appendix.

\section{Preliminaries and Problem Formulation}

\label{sec:pre}

Throughout the paper, the imaginary unit is denoted by $J:=\sqrt{-1}$. The set of integers $\{1,2,\ldots,k\}$ is denoted by $[k]$, $\mathbb{N}$ is the set of all positive integers, $\mathbb{R}$ is the set of all real numbers, and any other set is represented by a calligraphic capital letter. The cardinality of a set $\mathcal{X}$ is denoted
by $|\mathcal{X}|$. The difference of two sets $\mathcal{X}$ and $\mathcal{Y}$ is  denoted by $\mathcal{X} \fgebackslash \mathcal{Y}$. Matrices are represented by capital letters and vectors are expressed by boldface lower-case letters. Moreover, $\mathds{1}_{k}$ and $\mathbf{0}_{k}$ are $k$-dimensional column vectors with all ones and all zeros, respectively; and $\mathbf{e}_{i,k}$ is the $i$-th vector in the standard basis of $\mathbb{R}^{k}$. For a matrix $A$, $a_{ij}:=[A]_{ij}$ denotes its $ij$-th entry, and for a
block matrix $M$, $M_{(i)(j)}$ denotes the matrix corresponding
to the $ij$-th block. The $k\times k$ identity matrix is denoted by $I_k$, and $Z_k$ is the $k\times k$ matrix of all zeros.

\subsection{Algebraic Graph Theory}

A directed graph or \emph{digraph} $\mathcal{G}$ is defined as an
ordered pair of sets $\mathcal{G}:=(\mathcal{V},\mathcal{E})$, where
$\mathcal{V}=\left\{ \nu_{1},\ldots,\nu_{N}\right\} $ is a set of
$N=|\mathcal{V}|$ vertices (also called nodes or agents) and $\mathcal{E}\subseteq\mathcal{V}\times\mathcal{V}$
is a set of directed edges (also called links or arcs). Each edge
$\epsilon:=(\tau,\nu)\in\mathcal{E}$ is graphically represented by
a directed arc from vertex $\tau\in\mathcal{V}$ to vertex $\nu\in\mathcal{V}$.
Vertices $\nu$ and $\tau$ are referred to as the \emph{head} and
\emph{tail} of the edge $\epsilon$, and a $(\nu,\nu)$ edge is called
a self-loop on $\nu$. In our graphical representation, we
do not allow parallel edges, also called multi-edges, and we also assume that
graphs do not contain self-loops.

Given an integer $k\in\mathbb{N}$, an ordered set of (possibly repeated)
indices $\left(\alpha_{1},\alpha_{2},\ldots,\alpha_{k}\right)$ with
$\alpha_{i}\in$ $[N]$, and two vertices $\tau,\nu$ $\in$
$\mathcal{V}$, a \emph{$\tau\nu$-walk} of \emph{length} $k+1$ is
defined as an ordered sequence of directed edges of the form $\mathcal{W}$
$:=$ $((\tau,\nu_{\alpha_{1}}),(\nu_{\alpha_{1}},\nu_{\alpha_{2}}),$
$\ldots$ $,$ $(\nu_{\alpha_{k-1}},\nu_{\alpha_{k}}),(\nu_{\alpha_{k}},\nu))$.
A cycle or closed walk on node $\nu$ signifies a $\nu\nu$-walk.
For any $q,p\in[N]$, $\Omega^{k}(\nu_{q},\nu_{p})$ is
the set of all $\nu_{q}\nu_{p}$-walks in $\mathcal{G}$ with length
$k$. In the same venue, we define the distance
from $\nu_{q}$ to $\nu_{p}$ in $\mathcal{G}$ as 
\begin{align}
\textrm{dist}(\nu_{q},\nu_{p})=\min_{k\in\mathbb{N},\Omega^{k}(\nu_{q},\nu_{p})\neq\varnothing}k,\label{eq:distance}
\end{align}
where, by convention, $\textrm{dist}(\nu_{q},\nu_{q})=0$; and $\textrm{dist}(\nu_{q},\nu_{p})=\infty$
if $\Omega^{k}(\nu_{q},\nu_{p})=\varnothing$, for all $k\in\mathbb{N}$.
The diameter of $\mathcal{G}$, denoted by $\mbox{diam}(\mathcal{G})$,
is defined as the maximum distance between any pair of nodes $\nu_{p},\nu_{q}\in\mathcal{V}$: $\mbox{diam}(\mathcal{G}) = \max_{\nu_{q},\nu_{p} \in\mathcal{V}} \textrm{dist}(\nu_{q},\nu_{p})$.

The adjacency matrix of $\mathcal{G}$, which we denote by $G=\left[g_{ij}\right]$, is a binary matrix $G\in\{0,1\}^{N\times N}$ such that $G_{pq}=0$ for all pairs $p,q$ such that $(\nu_{q},\nu_{p})\not\in\mathcal{E}$ and $G_{pq}=1$, otherwise. The following is a well-known result from algebraic graph theory \cite{BiggsGraphTheory,Preciado:2013:MSA:2502376.2502379}:
\begin{mylemma}[Weighted Walk Counting]\label{lem:numWalksLemma}  Given the adjacency matrix $G$ of a directed
graph, the following identities hold: 
\begin{enumerate}
\item $\left[G^{k}\right]_{pq}=0,\mbox{ for all }k<\textrm{\emph{dist}}(\nu_{q},\nu_{p}),$ 
\item $\left[G^{k}\right]_{pq}>0\mbox{ for }k=\textrm{\emph{dist}}(\nu_{q},\nu_{p})$.
\label{eq:numWalksDistance} 
\end{enumerate}
\end{mylemma}

\subsection{Matrices and their Kronecker Algebra}

For matrices $A\in\mathbb{R}^{n\times m}$ and $B\in\mathbb{R}^{p\times q}$,
their Kronecker product $A\otimes B$ is defined as: 
\begin{align}
A\otimes B=\begin{bmatrix}a_{11}\mathbf{B} & \cdots & a_{1n}\mathbf{B}\\
\vdots & \ddots & \vdots\\
a_{m1}\mathbf{B} & \cdots & a_{mn}\mathbf{B}
\end{bmatrix},\label{eq:kronecker}
\end{align}
and given matrices $M_{1},M_{2},M_{3}$ and $M_{4}$ of appropriate
dimensions, the following identities are always satisfied: 
\begin{align}
(M_{1}\otimes M_{2})(M_{3}\otimes M_{4}) & =(M_{1}M_{3})\otimes(M_{2}M_{4}),\label{eq:kroneckerMIxedProd}\\
(M_{1}\otimes M_{2})\otimes M_{3} & =M_{1}\otimes(M_{2}\otimes M_{3}),\\
(M_{1}\otimes M_{2})^{-1} & =M_{1}^{-1}\otimes M_{2}^{-1}.\label{eq:kroneckerInverse}
\end{align}
In particular, for any $k\in\mathbb{N}$, \eqref{eq:kroneckerMIxedProd}
implies $(M_{1}\otimes M_{2})^{k}=M_{1}^{k}\otimes M_{2}^{k}$.

\subsection{Network Dynamic Model} \label{sec:formulation}

Consider a network of $N$ identical LTI subsystems whose interaction
structure is expressed by a directed information flow graph $\mathcal{G}=(\mathcal{V},\mathcal{E})$ with adjacency matrix $G=\left[g_{ij}\right]$. Let $\mathbf{x}_{(i)}(t)\in\mathbb{R}^{d}$ be the state of the system in node $\nu_{i}$ at time $t$. The evolution of this subsystem for time $t>t_{0}\in\mathbb{R}$ is given by: 
\begin{align}
\dot{\mathbf{x}}_{(i)}\left(t\right)= & A\mathbf{x}_{(i)}\left(t\right)+B\left(\sum_{q=1}^{N}g_{iq}\Gamma\mathbf{y}_{(q)}\left(t\right)+\mathbf{w}_{(i)}\left(t\right)\right),\label{stateEquations}\\
\mathbf{y}_{(i)}\left(t\right)= & C\mathbf{x}_{(i)}\left(t\right),\label{outputEquations}
\end{align}
where $\mathbf{y}_{(i)}\left(t\right)\in\mathbb{R}^{o}$ is the output
of the $i-$th subsystem. Similarly, $\mathbf{w}_{(i)}\left(t\right)\in\mathbb{R}^{m}$
is a vector of exogenous input signals injected into the $i-$th subsystem.
Matrices $A\in\mathbb{R}^{d\times d}$, $B\in\mathbb{R}^{d\times m}$
and $C\in\mathbb{R}^{o\times d}$ describe the evolution of each subsystem
in isolation, and $\Gamma\in\mathbb{R}^{m\times o}$ is called the
inner-coupling matrix describing how the output of neighboring nodes
influence the state of the $i$-th subsystem. We can describe the
global network dynamics using a set of `stacked' vectors, $\mathbf{x}:=(\mathbf{x}_{(1)}^{T},\ldots,\mathbf{x}_{(N)}^{T})^T$,
$\mathbf{y}:=(\mathbf{y}_{(1)}^{T},\ldots,\mathbf{y}_{(N)}^{T})^{T}$ and $\mathbf{w}:=(\mathbf{w}_{(1)},\ldots,\mathbf{w}_{(N)}^{T})^{T}$, and Kronecker products, as follows, 
\begin{align}
\dot{\mathbf{x}}\left(t\right) & =(I_{N}\otimes A+G\otimes B\Gamma C)\mathbf{x}\left(t\right)+\left(I_{N}\otimes B\right)\mathbf{w}\left(t\right), \nonumber\\
\mathbf{y}\left(t\right) & =\left(I_{N}\otimes C\right)\mathbf{x}\left(t\right).\label{eq:StateSPACEFailedSystem}
\end{align}

Part of our analysis will be performed in frequency domain. In this
context, each individual subsystem can be represented as a $o\times m$
transfer matrix $H(s)=C(sI_d-A)^{-1}B$. We denote by $r$ the least
relative degree among all the entries of $H(s)$, i.e., the least
difference between the degrees of the polynomials in the denominator
and the numerator of each entry. This quantity will be important in
the succeeding derivations.

\subsection{Detection, Isolation, and Sensor Location Problems}

Let us now state the particular problems considered in this paper.
We describe the \emph{detection problem} first. We assume that a central
entity, which we call `\emph{designer}', has access to the outputs
of a subset of nodes $S\subseteq\mathcal{V}$. We also assume that
the designer knows the nominal network information flow digraph $\mathcal{G}$
(the `faultless' graph). Neither the location of the failure, nor
the time of failure $t_{f}$ are known by the designer. In the
\emph{failure detection problem}, the designer is interested in determining
the existence of a single link failure, irrespective of its location,
at any given time. In the \emph{failure isolation problem}, however,
the designer would like to determine, not only the existence of a
failure, but also its location.

In this paper, we propose a detection and isolation algorithm based
on the analysis of abrupt changes in the derivatives, up to a certain
order $z$, of the sensor outputs induced by the failure, i.e., the designer
has access to the derivatives, $\left.d^{k}\mathbf{y}_{p}\right/dt^{k}$
for $k=1,\ldots,z$ and $\nu_{p}\in S$.

In the \emph{sensor location problem}, the designer needs to choose the location of a set
of sensors $S\subseteq\mathcal{V}$ to be able to detect and isolate
any potential edge failure for a given graph $\mathcal{G}$. The optimal
solution of the problem is achieved when the designer solves the problem
using the minimum number of sensors $\left|S\right|$. This problem
is combinatorial in nature and can be shown to be NP-hard, by reductions to the set-cover problems \cite{chvatal1979setcover}.
In this paper, we propose a greedy algorithm to approximate the optimal
solution to this problem and provide quality guarantees using submodular
set functions. In particular, we prove that our algorithm is a $(\log\left|\mathcal{E}\right|)+1$
approximation of the optimal combinatorial problem, in the sense that the cardinality of the sensor set returned by the proposed algorithm is no more than $(\log\left|\mathcal{E}\right|)+1$ times the minimum number of sensors required for the detection and isolation tasks.

\section{Failure Detection and Isolation}\label{sec:main}

In this Section we propose a methodology to detect and isolate link
failures in the dynamic network model proposed in Subsection \ref{sec:formulation}.
Our algorithm is based on the analysis of the derivatives of the output
of sensor nodes $S$ induced by a link failure. The impact on the
dynamics of a particular link failure can be replicated by a carefully
designed exogenous input, which we call the \emph{fault-replicant input} and denote by $\mathbf{f}(t)$. In the rest of the section, we first analyze the effect of the fault-replicant input on the sensor measurements, in particular, on its derivatives
(Subsection \ref{sub:ModelingFaults}). In Theorem \ref{theo:detection1},
we provide an exact characterization of the effect of a faulty link
on the sensor derivatives as a function of the distance, in number
of hops, from the faulty link to the sensor node. Based on this characterization,  in Subsection \ref{sub:FDI Algorithm}
we propose efficient algorithms to detect and isolate failures from the sensor outputs.

\subsection{Modeling a Single-Link Failure}\label{sub:ModelingFaults}

We start our analysis by considering the failure of a single link
$\bar{\epsilon}=(\nu_{j},\nu_{i})\in\mathcal{E}$ at time $t_{f}>t_{0}$.
Consequently, the information flow graph for $t>t_{f}$ is given by
$\bar{\mathcal{G}}=(\mathcal{V},\mathcal{E}\fgebackslash\{\bar{\epsilon}\})$,
which implies that $g_{ij}=0$ for $t>t_{f}$. Hence, the dynamics
of the $i$-th subsystem (the `head' of the faulty directed link $\bar{\epsilon}$)
for $t>t_{0}$ is given by, 
\begin{align}
\dot{\mathbf{x}}_{(i)}\left(t\right)= & A\mathbf{x}_{(i)}\left(t\right)+B\left(\sum_{q\neq j}g_{iq}\Gamma\mathbf{y}_{(q)}\left(t\right)+\mathbf{w}_{(i)}\left(t\right)\right).\label{eq:FaultyDynamics11}
\end{align}
It is convenient to replicate the dynamics of the network after link
$(\nu_{j},\nu_{i})$ fails by injecting an exogenous input $\mathbf{f}_{(i)}\left(t\right)$
into the $i$-th node of the \emph{faultless} network, as follows,
\begin{align}
\dot{\mathbf{x}}_{(i)}\left(t\right)=&A\mathbf{x}_{(i)}\left(t\right) \label{eq:dynamics} \\
& + B\left(\sum_{q=1}^{N}g_{iq}\Gamma C\mathbf{x}_{(q)}\left(t\right)+\mathbf{f}_{(i)}(t)+\mathbf{w}_{(i)}(t)\right),\nonumber 
\end{align} where the exogenous, fault-replicant input is defined as 
\begin{align}
\mathbf{f}_{(i)}\left(t\right):=\left\{ \begin{array}{cc}
0, & \mbox{for \ensuremath{t<t_{f}},}\\
-g_{ij}\Gamma C\mathbf{x}_{(j)}\left(t\right), & \mbox{for }t\geqslant t_{f}.
\end{array}\right.\label{eq:ReplicantInput}
\end{align}
Notice that using $\mathbf{f}_{(i)}\left(t\right)+\mathbf{w}_{(i)}\left(t\right)$
defined above as the exogenous input for the $i$-th subsystem results
in the faulty dynamics in \eqref{eq:FaultyDynamics11}. We can incorporate
the fault-replicant input into the faultless, global network dynamics
in \eqref{eq:StateSPACEFailedSystem}, by adding an exogenous input $\mathbf{f}\left(t\right):=\mathbf{e}_{i,N}\otimes\mathbf{f}_{(i)}\left(t\right)$.
Since, $\mathbf{x}_{(j)}\left(t\right)=\left(\mathbf{e}_{j,N}^{T}\otimes I_{d}\right)\mathbf{x}\left(t\right)$, we have, from (\ref{eq:ReplicantInput}), 
\begin{align}
\mathbf{f}\left(t\right)= & \mathbf{e}_{i,N}\otimes\left(-g_{ij}\Gamma C\left(\mathbf{e}_{j,N}^{T}\otimes I_{d}\right)\mathbf{x}\left(t\right)\right)\nonumber \\
= & -g_{ij}\left(\mathbf{e}_{i,N}\mathbf{e}_{j,N}^{T}\otimes\Gamma C\right)\mathbf{x}\left(t\right),\label{fkron}
\end{align} for $t\geqslant t_{f}$; and $\mathbf{f}\left(t\right)=\mathbf{0}_{Nm}$ for $t<t_{f}$.

\subsection{Impact of Single-Link Failures on Output Derivatives}

We now proceed to analyze the effect of the fault-replicant input
on the sensor measurements, in particular, on their derivatives. We
present a theorem that characterizes the effect of single-link faults
on the derivatives of the sensor outputs as a function of the distance
from the faulty link to the sensor node. We state the theorem in terms
of the following function, 
\begin{equation}
\mathbf{\Delta}_{p,k}(t):=\lim_{\varepsilon\to0^{+}}\left(\left.\frac{d^{k}\mathbf{y}_{(p)}}{dt^{k}}\right|_{t+\varepsilon}-\left.\frac{d^{k}\mathbf{y}_{(p)}}{dt^{k}}\right|_{t-\varepsilon}\right),\label{eq:DerivativeJump}
\end{equation}
which measures the jump in the $k$-th derivative of the output of node $p$ at time $t$: $\mathbf{y}_{(p)}(t)$.

\begin{mytheorem}[Jump Discontinuities of Output Derivatives]\label{theo:detection1} Consider the dynamic network in \eqref{stateEquations}-\eqref{outputEquations} where the exogenous input $\mathbf{w}_{(i)}$ is $(\mbox{\emph{diam}}\left(\mathcal{G}\right)+1)r$-differentiable and $r$ is the least relative degree among all the entries of the
transfer matrix $H(s)=C(sI_d-A)^{-1}B$. Assume that link $\bar{\epsilon}=(\nu_{j},\nu_{i})$
fails at time $t_{f}$. Then, $\mathbf{\Delta}_{p,k}(t_{f}) = \mathbf{0}_{o}$ for $k<(\mbox{\emph{dist}}(\nu_{i},\nu_{p})+1)r$ and {$\mathbf{\Delta}_{p,k}(t_{f}) = g_{ij}\left[G^{k}\right]_{pi} Q \Gamma C\mathbf{{x}}_{(j)}(t_f)$}, if $k=(\mbox{\emph{dist}}(\nu_{i},\nu_{p})+1)r$, where all distances are calculated w.r.t. the original digraph $\mathcal{G}$, and $Q \in \mathbb{R}^{o\times o}$ is a constant matrix given by {$Q  := \mbox{lim}_{s\to\infty}s^{k}\left[H(s)\Gamma\right]^{\textrm{\emph{dist}}(\nu_{i},\nu_{p})+1}$}.
\end{mytheorem}

\begin{myremark} The above theorem characterizes the effect of a fault in link $(\nu_{j},\nu_{i})$ on the $k$-th derivative of the output of node $\nu_{p}$ as a function of the distance (in number of hops) from the head of the faulty link $\nu_{i}$ to the output node $\nu_{p}$, and the relative degree $r$ of the transfer matrix $H(s)$.
\end{myremark}

\subsection{Algorithms for Fault Detection and Isolation\label{sub:FDI Algorithm}}

In this subsection, we propose an algorithm for fault detection and
isolation based on the preceding results. Theorem~\ref{theo:detection1}
states that the failure of link $(\nu_{j},\nu_{i})$ induces a jump
in the $k$-th derivative of the output of node $\nu_{p}$ for $k=(\mbox{dist}(\nu_{i},\nu_{p})+1)r$.
We can therefore design a simple detection algorithm by monitoring the
jumps in the derivatives of the sensor nodes. Specifically, if $\mathbf{\Delta}_{p,k}(t_{f})\neq\mathbf{0}_{o}$ for any $\nu_{p}\in S$, $k\leq z$ and $t_{f}>t_{0}$, then a link
has failed in the network at time $t_{f}$. The following condition
for detectability of a link failure is a direct consequence of Theorem
\ref{theo:detection1}:

\begin{mycorollary}[Detectable Links]\label{coro:detectableLinks}The failure of link $\left(\nu_{j},\nu_{i}\right)$
is detectable using the set of output derivatives $\{\frac{d^{k}}{dt^{k}}\mathbf{y}_{(p)}:\nu_{p}\in S,k\leq z\}$
if there exists a directed path of length $l\leq\left(\frac{z}{r}\right)-1$
from $\nu_{i}$ to a node in $S$.\end{mycorollary}

Once a link failure is detected, it may also be isolated by looking
at the values of $p$ and $k$ for which $\mathbf{\Delta}_{p,k}(t_{f}) \neq \mathbf{0}_{o}$,
under certain condition on the distribution of sensors, which we study in Section \ref{sec:SensorPlacement}. Our strategy
is based on exploiting the relationship between the location of the
failure and abrupt changes in the output derivatives of a particular
order. To achieve this, it is convenient to define a look-up table
$D$, with $d_{p\epsilon}=\left[D\right]_{p\epsilon}\in\left\{ 0,1,\ldots,z\right\} ^{\left|S\right|\times\left|\mathcal{E}\right|}$,
with columns indexed by edges and rows by sensor nodes. Given a communication
graph $\mathcal{G}$, the designer can compute the matrix $D$, running
a double loop search over the set of edges $\epsilon\in\mathcal{E}$ and
the set of sensor nodes $\nu_{p}\in S$. For each edge $\epsilon=\left(\nu_{j},\nu_{i}\right)$
and node $\nu_{p}$ in this loop, the designer assigns $d_{p\epsilon}:=(\mbox{dist}(\nu_{p},\nu_{i})+1)r$.
Although matrix $D$ is defined using distances in the communication
graph, its entries has a dynamical interpretation. According to Theorem
\ref{theo:detection1}, if link $\epsilon\in\mathcal{E}$ fails at
time $t_{f}$, the we have
\begin{align}
\min_{k\leq z}\left\{ k:\mathbf{\Delta}_{p,k}(t_{f})\neq \mathbf{0}_{o}\right\} =d_{p\epsilon}.
\end{align}
In other words, $d_{p\epsilon}$ is equal to the smallest order $k$
of the derivative of sensor $p$'s output $\mathbf{y}_{p}$, at which
we observe an abrupt change when link $\epsilon$ fails. Therefore,
using the matrix $D$, the designer may be able to isolate a link
failure, as follows. First, when a link failure is detected at $t=t_{f}$,
the designer constructs a column vector $\mathbf{k}=\left(k_{1},\ldots,k_{\left|S\right|}\right)^{T}$,
where $k_{p}:=\min_{k\leq z}\left\{ k:\mathbf{\Delta}_{p,k}(t_{f}) \neq \mathbf{0}_{o}\right\} $,
i.e., $k_{p}$ is the smallest order of the derivative at which the
designer observes a jump in the output of sensor node $\nu_{p}$.
Notice that, if the vector $\mathbf{k}$ matches one, and only one,
of the $\left|\mathcal{E}\right|$ columns of $D$, then the designer can isolate
the location of the link failure. In particular, if $\mathbf{k}$
matches the $\epsilon$-th column of $D$, then the designer can conclude
that link $\epsilon$ has failed. In Section \ref{sec:SensorPlacement},
we propose a sensor location algorithm to guarantee the existence
and uniqueness of a column in $D$ matching $\mathbf{k}$ for any
possible link failure. We illustrate the above methodology for failure
isolation in the following example.

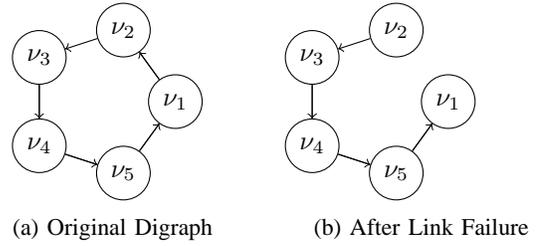
\begin{figure}
\centering
\begin{subfigure}[b]{0.2\textwidth}
\hspace*{10pt}
\begin{tikzpicture}
\tikzstyle{every node}=[draw,shape=circle];
\node (v0) at ( 0:1) {$\nu_1$};
\node (v1) at ( 72:1) {$\nu_2$};
\node (v2) at (2*72:1) {$\nu_3$};
\node (v3) at (3*72:1) {$\nu_4$};
\node (v4) at (4*72:1) {$\nu_5$};

\foreach \from/\to in {v0/v1, v1/v2, v2/v3, v3/v4, v4/v0}
\draw [->] (\from) -- (\to);

\draw
(v0) -- (v1)
(v2) -- (v3)
(v3) -- (v4)
(v4) -- (v0);
(v1) -- (v2);

\end{tikzpicture}
\caption{Original Digraph}
\end{subfigure}~\hspace*{10pt}\begin{subfigure}[b]{0.2\textwidth}
\begin{tikzpicture}
\tikzstyle{every node}=[draw,shape=circle];
\node (v0) at ( 0:1) {$\nu_1$};
\node (v1) at ( 72:1) {$\nu_2$};
\node (v2) at (2*72:1) {$\nu_3$};
\node (v3) at (3*72:1) {$\nu_4$};
\node (v4) at (4*72:1) {$\nu_5$};

\foreach \from/\to in { v1/v2, v2/v3, v3/v4, v4/v0}
\draw [->] (\from) -- (\to);

\draw

(v2) -- (v3)
(v3) -- (v4)
(v4) -- (v0);
(v1) -- (v2);

\end{tikzpicture}
\caption{After Link Failure}
\end{subfigure}
\caption{A Directed Cycle of Length Five}
\label{fig:dir-ring-graph}
\end{figure}

\begin{myexample}\label{example:5nodeCycle} Isolation of a Link Failure. \end{myexample}
Consider the cycle network in Fig~\ref{fig:dir-ring-graph}. Agents in this
network are single integrators following a Laplacian dynamics, i.e.,
$\dot{\mathbf{x}}=-\mathcal{L}(\mathcal{G})\mathbf{x}$, where $\mathcal{L}(\mathcal{G})$ is the Laplacian matrix
of the network. Assume that we have two sensor nodes, $S=\{\nu_{2},\nu_{3}\}$, and the edges are labeled such that the edge set is given by $\mathcal{E}$ $=$ $\{\epsilon_{q},q\in[5]\}$,
where for $q\in[5]\fgebackslash\{1\}$, $\epsilon_{q}=(\nu_{q-1},\nu_{q})$
and $\epsilon_{1}=(\nu_{5},\nu_{1})$. Following the methodology proposed in this subsection, we can construct
the following look-up table matrix, 
\[
D=\left(\begin{array}{ccccc}
2 & 1 & 0 & 4 & 3\\
3 & 2 & 1 & 0 & 4
\end{array}\right),
\]
where the entries in the $\epsilon$-th column of $D$, for $\epsilon=1,\ldots,5$,
are the distances, plus one, from the head of the $\epsilon$-th link
to each one of the nodes in $S=\{\nu_{2},\nu_{3}\}$. Let us assume, for example, a failure
in edge $\epsilon_{2} = (\nu_1,\nu_2)$. According to Theorem \ref{theo:detection1},
this failure induces abrupt changes in the $1^{st}$ and $2^{nd}$
derivatives of nodes $2$ and $3$, respectively, which results in
a vector $\mathbf{k}=\left(1,2\right)^{T}$. Notice that this vector
matches the second column of $D$; therefore, the designer is able
to detect and isolate this failure from changes in the first two derivatives
of the sensor nodes, by matching the order of derivatives in which the jumps are observed with the columns of $D$.

\section{Efficient Sensor Placement}\label{sec:SensorPlacement}

It is not always possible to isolate, or even detect, link failures
from a given set of sensor locations, $S$. In this section, we study
the problem of distributing sensors in a network to guarantee that
the designer is able to detect and isolate any possible link failure.
Although finding the minimum number of sensors to accomplish this
goal is a hard combinatorial problem,
we provide in this section efficient approximation algorithm with
quality guarantees based on submodular set functions and the set-cover problem. In particular,
we prove that our algorithm is a $\log\left|\mathcal{E}\right|+1$
approximation to the optimal combinatorial problem. In other words,
the size of the sensor set $S$ rendered by the approximation algorithm
is at most $\log\left|\mathcal{E}\right|+1$ times the minimum number
of sensors required for fault detection and isolation.

Before describing our sensor placement algorithm, we define a set of
binary relations that will be used later on.
{
\begin{mydef}[Binary Relations between Edges and Nodes]\label{Def:BinaryRelationships}We define the following binary relations
between $[N]$ and $\mathcal{E}$, denoted by $\mathcal{R}_{0}$
and $\mathcal{R}_{k}$, for $k\in[z]$. For all $p\in[N]$ and $\epsilon=(\nu_{q},\nu_{s})\in\mathcal{E}$:
\begin{itemize}
\item $(p,\epsilon)\in\mathcal{R}_{k}\iff k=(\textrm{\emph{dist}}(\nu_{s},\nu_{p})+1)r$,
\item $(p,\epsilon)\in\mathcal{R}_{0}\iff({p},\epsilon)\not\in\mathcal{R}_{k}$
for any $k\in[z]$.
\end{itemize}
\end{mydef} 
\begin{myremark} $\mathcal{R}_{k}$ is defined such that if $(p,\epsilon)\in\mathcal{R}_{k}$,
then the failure of link $\epsilon$ produces a jump in the $k$-th
derivative of the response of node $p$. On the other hand, if $(p,\epsilon)\in\mathcal{R}_{0}$,
then the failure of edge $\epsilon$ does not produce a jump in any
of the derivatives of the response of node $p$ up to the $z$-th
order.
\end{myremark}}
Based on the above definition, we propose in the following subsections
efficient algorithms for sensor placement. We also provide quality
guarantees based on the results developed in the context of set-cover
problems \cite{Dhillon03sensorplacement,Leskovec}.

\subsection{Detection of Link Failures: Coverage\label{sec:coverage}}

The link-failure detection problem can be restated using the binary
relation $\mathcal{R}_{0}$, as follows:

\begin{myproblem}[Detection]\label{prob:detection}Given a digraph $\mathcal{G}=(\mathcal{V},\mathcal{E})$,
find a subset of nodes $\mathcal{M}_{D}\subseteq\mathcal{V}$ of minimum
cardinality $\left|\mathcal{M}_{D}\right|$, such that for all $\epsilon\in\mathcal{E}$,
there exists a node $\nu_{p}\in\mathcal{M}_{D}$ with $({p},\epsilon)\not\in\mathcal{R}_{0}$.
\end{myproblem}

Let $\mathcal{M}_{D}$ be any solution to the above problem. Then,
the failure of any link at time $t_{f}$ would induce an abrupt change
in some of the first $z$ derivatives of the output of at least one
node in $\mathcal{M}_{D}$. In other words, by observing the first
$z$ derivatives of the outputs of all nodes in $\mathcal{M}_{D}$,
one is able to determine that a failure occurs. Finding $\mathcal{M}_{D}$
is a hard combinatorial problem for which we propose an approximation
algorithm below. To present this approximation, it is convenient to
define the following concepts:

\begin{mydef}[Submodular Functions]\label{def:submodular}
Given two sets $\hat{\mathcal{M}},\bar{\mathcal{M}}\subset\mathcal{V}$
such that $\hat{\mathcal{M}}\subset\bar{\mathcal{M}}$, a function\footnote{Given a set $\mathcal{X}$, we denote by $\mathcal{P}(\mathcal{X})=\{\mathcal{M} \mathcal{M}\subset\mathcal{X}\}$ the power-set of $\mathcal{X}$, which is the set of all its subsets.} $f:\mathcal{P}\left(\mathcal{V}\right)\to\mathbb{R}_{+}$ is submodular
if for all $\nu_{q}\in\mathcal{V}$ the following inequality holds true $f(\hat{\mathcal{M}}\cup\{\nu_{q}\})-f(\hat{\mathcal{M}})\leqslant f(\bar{\mathcal{M}}\cup\{\nu_{q}\})-f(\bar{\mathcal{M}})$. \end{mydef}

\begin{mydef}[Coverage Function]\label{def:coverageFunction} Given a set of nodes $\mathcal{M}\subset\mathcal{V}$, we define the ``\emph{coverage function}'' $f_{D}:\mathcal{P}(\mathcal{V})\to\left[|\mathcal{E}|\right]\cup\{0\}$, as follows $f_{D}(\mathcal{M})=|\{\epsilon\in\mathcal{E}:\forall p\in\mathcal{M},(p,\epsilon)\in\mathcal{R}_{0}\}|$. In other words, this is the number of edges in the network whose failure would \emph{not} induce a jump in any of the first $z$ derivatives of the outputs of any node in $\mathcal{M}$. \end{mydef}

\begin{claim}[Submodularity of Coverage Function]\label{claim:f_D} {The function} $-f_{D}(\cdot)$ is a submodular set function from $\mathcal{P}(\mathcal{V})$
to $\left[|\mathcal{E}|\right]\cup\{0\}$.\end{claim}

In what follows, we propose a greedy algorithm \cite{Krause} to efficiently
find an approximate solution $M_{D}$ to Problem \ref{prob:detection}.
We provide performance guarantees in Subsection \ref{sec:setcover}.

\begin{algorithm}
\caption{Determine a Solution $\mathcal{M}_D$ to Problem \ref{prob:detection}}
\label{routine:detection}
\begin{algorithmic}[1]
\REQUIRE $\mathcal{G} = (\mathcal{V},\mathcal{E})$
\State $\mathcal{M}_D \Leftarrow \varnothing$
\WHILE{$f_D(\mathcal{M}_D) \neq 0$}
\STATE $\nu_q  \Leftarrow \arg\min \{ f_D({\mathcal{M}}_D\cup\{\nu_q\}) - f_D({\mathcal{M}}_D);\nu_q \in \mathcal{V}\fgebackslash {\mathcal{M}}_D\}$
\STATE ${\mathcal{M}}_D \Leftarrow {\mathcal{M}}_D\cup\{\nu_q\}$
\ENDWHILE
\ENSURE ${\mathcal{M}}_D$
\end{algorithmic}
\end{algorithm}

\begin{myremark}\label{rem:rout1} The function $f_{D}(\mathcal{M}_{D})$
measures the coverage of set $\mathcal{M}_{D}$ by counting the number
of links that are not yet covered by $\mathcal{M}_{D}$. At each iteration
of Routine \ref{routine:detection}, the extra agent $q$ is selected
and added to $\mathcal{M}_{D}$ such that the number of newly covered
links is maximized. Note that since for any $\epsilon=(\nu_{q},\nu_{p})\in\mathcal{E}$,
$(p,\epsilon)\in\mathcal{R}_{1}$ it follows that $f_{D}(\mathcal{V})=0$,
whence Routine \ref{routine:detection} is guaranteed to terminate.
\end{myremark}

The focus is next shifted to the isolation problem, for which a similar
routine is developed.

\subsection{Isolation of Link Failures: Resolution}\label{sec:resolution}

For analyzing the isolation problem, we introduce several definitions
that will be useful to describe our isolation algorithm.
\begin{mydef}[Indicator Set of an Edge]\label{def:IndicatorSet} Given a subset of agents $\mathcal{M}\subseteq\mathcal{V}$ and an edge $\epsilon\in\mathcal{E}$, we define the ``\emph{indicator set}'' of $\epsilon$ w.r.t. $\mathcal{M}$ as the correspondence $\mathcal{I}:\mathcal{P}(\mathcal{V})\times\mathcal{E}\to\mathcal{P}(([z]\cup\{0\})\times\mathcal{V})$, given by $\mathcal{I}(\mathcal{M},\epsilon)=\{(k,\nu_{p})\in([z]\cup\{0\})\times\mathcal{M})\mbox{ such that }(p,\epsilon)\in\mathcal{R}_{k}\}$. In other words, given an edge $\varepsilon$ and a set of nodes $\mathcal{M}$, this indicator set is the set of all the pairs $(k,\nu_{p})$ where
$k$ is the order of the derivative at which there is a jump in the output of node $\nu_{p}\in\mathcal{M}$ when edge $\varepsilon$ fails.
\end{mydef}

\begin{mydef}[Resolution Function]\label{def:ResolutionFunction} Given a subset of agents $\mathcal{M}\subseteq\mathcal{V}$, define the set of \emph{unidentified edges} associated with $\mathcal{M}$ as the correspondence $\mathcal{U}:\mathcal{P}(\mathcal{V})\to\mathcal{P}(\mathcal{E})$, given by $\mathcal{U}(\mathcal{M}) = \{\epsilon\in\mathcal{E}:\exists\hat{\epsilon}\in\mathcal{E}\fgebackslash\{\epsilon\}\mbox{ \emph{for which} }\mathcal{I}(\mathcal{M},\epsilon)=\mathcal{I}(\mathcal{M},\hat{\epsilon})\}$. Defined such, $\mathcal{U}(\mathcal{M})$ is the set of all edges whose failures are not uniquely identified based on the order of the derivative at which there is a jump in the output responses of the nodes in $\mathcal{M}$. We further define the ``\emph{resolution function}'', $f_{I}(\mathcal{M})=|\mathcal{U}(\mathcal{M})|$, as the number of edges that are not uniquely identified based on the orders of the derivatives at which we observe jumps in the outputs of the nodes in $\mathcal{M}$.
\end{mydef}

\begin{claim}[Submodularity of Resolution Function]\label{claim:f_I}
{The function} $-f_{I}(\cdot)$ is a submodular set function from $\mathcal{P}(\mathcal{V})$
to $\left[|\mathcal{E}|\right]\cup\{0\}$. \end{claim}

Next using the above concepts, we can restate the isolation problem,
as follows:

\begin{myproblem}[Isolation]\label{prob:isolation} Given a digraph $\mathcal{G}=(\mathcal{V},\mathcal{E})$, find a subset of vertices $\mathcal{M}_{I}\subset\mathcal{V}$ with the smallest cardinality $\left|\mathcal{M}_{I}\right|$, such that $f_{I}\left(\mathcal{M}_{I}\right)=0$.
\end{myproblem}
\begin{myremark}
Note that when $f_{I}\left(\mathcal{M}_{I}\right)=0$, there are no
edges in the network that cannot be uniquely identified based on the
jumps in the derivatives of the outputs of nodes in $\mathcal{M}_{I}$.
In other words, all edge failures can be isolated.\end{myremark}
\begin{mycorollary}
Isolation of all edge failures is possible if and only if $f_{I}\left(\mathcal{V}\right)=0$.
\end{mycorollary}
Since Problem \ref{prob:isolation} is hard to solve exactly, we now
propose a greedy heuristic similar to the one in Subsection \ref{sec:coverage}
to find an approximate solution, as follows. We also provide quality
guarantees in Subsection \ref{sec:setcover}.

\begin{algorithm}
\caption{Determine a Solution $\mathcal{M}_I$ to Problem \ref{prob:isolation}}
\label{routine:isolation}
\begin{algorithmic}[1]
\REQUIRE $\mathcal{G} = (\mathcal{V},\mathcal{E}) \And \mathcal{M}_D$
\State $\mathcal{M}_I \Leftarrow \mathcal{M}_D$
\WHILE {$f_I(\mathcal{M}_I) \neq 0 \And \mathcal{M}_I \neq \mathcal{V}$}
\STATE $\nu_q  \Leftarrow \arg\min \{ f_I({\mathcal{M}}_I\cup\{\nu_q\}) - f_I({\mathcal{M}}_I);\nu_q \in \mathcal{V}\fgebackslash {\mathcal{M}}_I\}$
\STATE ${\mathcal{M}}_I \Leftarrow {\mathcal{M}}_I\cup\{\nu_q\}$
\ENDWHILE
\IF{$f_I(\mathcal{M}_I) \neq 0$} \State $\mathcal{M}_I \Leftarrow \varnothing$
\ENDIF 
\ENSURE ${\mathcal{M}}_I$
\end{algorithmic}
\end{algorithm}

\begin{myremark}\label{rem:rout2_remark} Note that a solution $\mathcal{M}_{D}$ to Problem~\ref{prob:detection}
is required as an input for Routine~\ref{routine:isolation}, with
which $\mathcal{M}_{I}$ is initialized. This ensures that any valid
output of Routine~\ref{routine:isolation} also satisfies the detection
requirements, and such an initialization is required because the $\mathcal{R}_{0}$
relations (lack of jumps in some nodes) are just as informative for
the isolation purposes. In other words, unlike the detection problem
where the goal is to ensure a non-zero relation at one or more observation
points for any edge in the network; in the case of isolation, a link
may just as well be identified through its $\mathcal{R}_{0}$ relations
with all the observation points. This, however, renders the failure
of the link in question undetectable and it is exactly to prevent
such a case that the initialization step in Routine~\ref{routine:isolation}
is necessary.\end{myremark}

\begin{myremark}\label{rem:rout2} The function $f_{I}(\mathcal{M}_{I})$
measures the resolution of set $\mathcal{M}_{I}$ by counting the
number of links that are not uniquely identified through their relations
$\mathcal{R}_{k}$ with the vertexes of set $\mathcal{M}_{I}$. At
each iteration of Routine \ref{routine:isolation}, the extra agent
$q$ is selected and added to $\mathcal{M}_{I}$ such that the resultant
improvement in the resolution of $\mathcal{M}_{I}$ is maximized.
Note that unlike Problem \ref{prob:detection}, it is possible for
Problem \ref{prob:isolation} to have no solutions at all, in which
case Routine \ref{routine:isolation} returns $\varnothing$. This
occurs if, and only if, $f_{I}(\mathcal{V})\neq 0$. \end{myremark}

{

\subsection{Performance Guarantees from the Set Covering Problem}

\label{sec:setcover}

The following definition is most useful when implementing Routines~\ref{routine:detection}
and \ref{routine:isolation} in general networks and selecting possible
solutions $\mathcal{M}_{D}$ and $\mathcal{M}_{I}$ for Problems \ref{prob:detection}
and \ref{prob:isolation}, respectively. Label the edges
of the network from $1$ to $l=|\mathcal{E}|$
and denote them by $\mathcal{E}$ $=$ $\{\epsilon_{q},q\in[l]\}$.
Associate with every edge $\epsilon_{q},q\in[l]$ a row
vector $\mathbf{r}_{q}$ with $N$ columns, whose element $[\mathbf{r}_{q}]_{\gamma},\gamma\in[N]$
is equal to $k$ if $(\gamma,{\epsilon}_{q})\in\mathcal{R}_{k}$ for
some $k\in[z]\cup\{0\}$ and let matrix $R\in([z]\cup\{0\})^{l\times N}$
be the matrix whose $q-$th row is equal to $\mathbf{r}_{q}$ for
all $q\in[l]$. The highest order of derivatives $z$ that
is to be observed at the observation points is then set equal to the
maximum value of the entries of matrix $R$ and is thus bounded above
by the diameter of the network.

Problems \ref{prob:detection} and \ref{prob:isolation} can now be
restated in terms of the matrix $R$ as follows. For the detection
problem choose the columns of $R$ such that the corresponding entries
have at least one non-zero element per each row. By the same token,
for the isolation choose the columns such that the set of entries
ordered in the order of the chosen columns are not identical for any
two rows.

Let the location of the non-zero entries of the matrix ${R}$ of a
network be indexed by the one entries in a binary matrix $\tilde{R}$,
which has the same dimensions $l\times N$ as $R$. Problem \ref{prob:detection}
can then be formulated as finding $\mathbf{x}$ such that:

\begin{equation}
\begin{aligned} & \underset{\mathbf{x}\in{\{0,1\}}^{N}}{\text{minimize}} &  & \mathbf{\mathds{1}}_{N}^{T}\mathbf{x}\\
 & \text{subject to} &  & \tilde{R}\mathbf{x}\geqslant\mathbf{\mathds{1}}_{l}^{T}.
\end{aligned}
\label{eq:setCOver}
\end{equation}

The discrete optimization problem in \eqref{eq:setCOver} is an instance
of the set covering problem \cite{vohra1998}. Accordingly, given equal column weights, it is desired to choose a subset of columns
with minimum total weight, such that at least one non-zero element from each row is selected. For an integer $d\in\mathbb{N}$, define the truncated harmonic sum $\mathcal{H}(d)=\sum_{i=1}^{d}\frac{1}{i}$ and let $d_{max}$ be the maximum of the column sums associated with matrix $\tilde{R}$. It is known from the classical results \cite{chvatal1979setcover,Johnson1974,Lovasz1975,Dobson1982},
that the greedy heuristic in Routine \ref{routine:detection} is guaranteed
to produce a result ${\mathcal{M}}_{D}$, which is no worse than $\mathcal{H}(d_{max})$
times the value of the optimal answer $\mathbf{\mathds{1}}_{N}^{T}\mathbf{x}$
obtained from \ref{eq:setCOver}.

Similarly for Problem \ref{prob:isolation}, it lends itself to the
following formulation as a sub-modular set covering problem \cite{Wolsey1982},
\begin{equation}
\begin{aligned} & \underset{\mathcal{S}\subset[N]}{\text{minimize}} &  & \sum_{j\in\mathcal{S}}w_{j}\\
 & \text{subject to} &  & f_{I}(\{\nu_q,q\in\mathcal{S}\})=f_{I}(\mathcal{V}),
\end{aligned}
\label{eq:SubmodularSetCover}
\end{equation}
where using the unity column weights: $w_{j}=1,\forall j\in[N]$,
the objective can be rewritten as $\sum_{j\in\mathcal{S}}w_{j}=|\mathcal{S}|$.
It is again known from the classical theory that the value of the
solution of the greedy heuristic in Routine \ref{routine:isolation}
never exceeds the optimal value by more than a factor of $\mathcal{H}(\max_{j}\{l-f_{I}(\{\nu_j\})\})$,
\cite{Wolsey1982}. Noting the trivial bounds, $\mathcal{H}(d_{max})\leqslant H(|\mathcal{E}|)$,
$\mathcal{H}(\max_{j}\{l-f_{I}(\{\nu_j\})\})\leqslant \mathcal{H}(|\mathcal{E}|)$, and $\mathcal{H}(|\mathcal{E}|)\leqslant\log(|\mathcal{E}|)+1$,
it follows that the proposed sensor placement algorithms of Subsections
\ref{sec:coverage} and \ref{sec:resolution} provide answers that
are within a factor $\log(|\mathcal{E}|)+1$ of the optimal answer;
and therefore, they scale with a rate no worse than the logarithm
of the size of the network edge-set. }

\section{Examples and Discussions}\label{sec:examples}

In this section we provide additional examples and discussions of special cases, followed by computer experiments with a random geometric graph.

\begin{myexample}Star Networks.\end{myexample} \label{subsec:star}

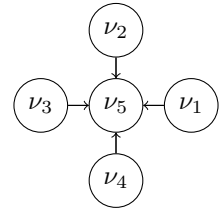
\begin{figure}
\centering
\begin{tikzpicture}
\tikzstyle{every node}=[draw,shape=circle];
\node (v0) at ( 0:0) {$\nu_5$};
\node (v1) at ( 0:1) {$\nu_1$};
\node (v2) at (1*90:1) {$\nu_2$};
\node (v3) at (2*90:1) {$\nu_3$};
\node (v4) at (3*90:1) {$\nu_4$};

\foreach \from/\to in {v0/v1, v0/v2, v0/v3, v0/v4}
\draw [<-] (\from) -- (\to);

\draw

(v0) -- (v1)
(v0) -- (v2)
(v0) -- (v3)
(v0) -- (v4);

\end{tikzpicture}
\caption{A Star Digraph of Size Five}
\label{fig:dir-star-graph}
\end{figure}

For this and the next subsection, let $\mathcal{V}=\{\nu_{q},q\in[5]\}$
be a set of five vertexes. As the first example, consider the case
of a star network (Fig.~\ref{fig:dir-star-graph}), where there are four edges in the network
and all of them share the same head vertex $\nu_{5}$. In this case,
the designer can detect the failure of any single edge in the network
by observing the first derivative of the response of $x_{5}$; however,
there are no subset of nodes that can be observed for isolating the
failed edge. In fact $f_{I}([5])=4$ in the case of a star
network, and every edge of the network is in the same relations $\mathcal{R}_{1}$
with the node $\nu_{5}$ and $\mathcal{R}_{0}$ with the rest of the
nodes. Indeed, this is best seen through its $R$ matrix given as $R = \left(Z_4\,\mathds{1}_{4}\right)$.

The above generalizes for finite star networks of arbitrary size.
In particular, there is no solution to the isolation problem for any
star network and detection can be achieved by observing the first
derivative of the common head vertex.

\begin{myexample}\label{subsec:sec}{Laplacian Dynamics on Cycle Networks.}\end{myexample}

As the second example, consider the five node cycle in Fig.\ref{fig:dir-ring-graph} and Example~\ref{example:5nodeCycle}. The matrix $R$ for this network
is given by:

\[
R=\left(\begin{array}{rrrrr}
1 & 2 & 3 & 4 & 0\\
0 & 1 & 2 & 3 & 4\\
4 & 0 & 1 & 2 & 3\\
3 & 4 & 0 & 1 & 2\\
2 & 3 & 4 & 0 & 1
\end{array}\right).
\]

\begin{figure}[ht]
\centering \includegraphics[width=0.48\textwidth]{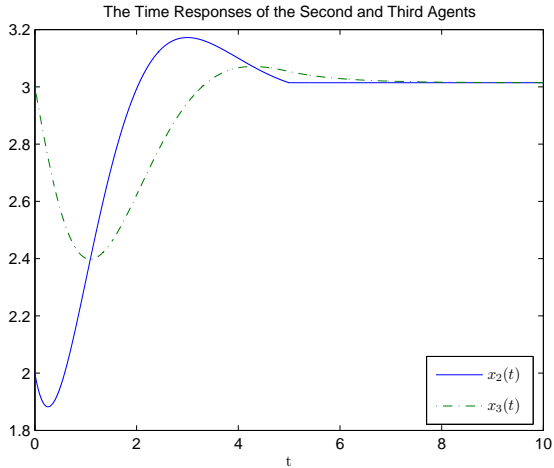} \caption{The output responses $x_{2}(t)$ and $x_{3}(t)$ are plotted for $0\leqslant t\leqslant10$.
The link failure happens at $t_{f}=5$, where there is a break in
the plot of $x_{2}(t)$ but not of $x_{3}(t)$.}

\label{fig:responses} 
\end{figure}

\begin{figure}[ht]
\centering \includegraphics[width=0.48\textwidth]{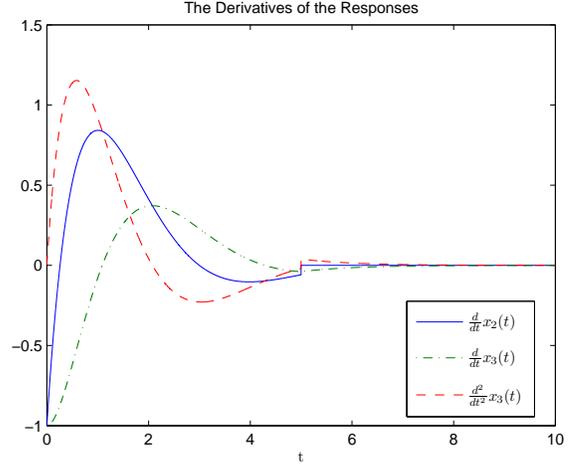} \caption{The derivatives of $x_{2}(t)$ and $x_{3}(t)$ are plotted for $0\leqslant t\leqslant10$.
At the time of failure $t_{f}=5$, there are jump discontinuities
in the plots of $\frac{\textrm{{d}}}{\textrm{{d}}t}x_{2}(t)$ and
$\frac{{\textrm{{d}}}^{2}}{\textrm{{d}}t^{2}}x_{3}(t)$. The plot
of $\frac{{\textrm{{d}}}^{2}}{\textrm{{d}}t^{2}}x_{2}(t)$ contains
an impulse at $t_{f}=5$, while $\frac{\textrm{{d}}}{\textrm{{d}}t}x_{3}(t)$
is a continuous function of time.}

\label{fig:derivatives} 
\end{figure}

It is evident from matrix $R$ that any two distinct vertexes offer
a solution $\mathcal{M}_{I}$ to Problem \ref{prob:detection}, since
the locations of $0$ entries do not overlap for distinct vertexes.
Thence, the designer can detect the failure of any links in the cycle
network by observing the jumps in the first four derivatives of any
two nodes in the network. In the same vein, any set of two distinct
vertexes can also be used to uniquely determine which link has failed
based on the observed jumps in the first four derivatives. For instance,
taking $\mathcal{M}_{I}=\{2,3\}$, $\epsilon_{2}$ is the only edge
whose failure produces a jump in the first and second derivatives
of $x_{2}(t)$ and $x_{3}(t)$, respectively, at the time of failure,
$t=t_{f}$. Fig \ref{fig:responses} depicts the responses of the
second and third agents, for a directed cycle of length five initialized
at $\mathbf{x}(0)=(1,2,3,4,5)^{T}$, where for all $t\in\mathbb{R}$,
$\mathbf{x}(t)=(x_{1}(t),x_{2}(t),x_{3}(t),x_{4}(t),x_{5}(t))^{T}$
and $\dot{\mathbf{x}}(t)=-\mathcal{L}(\mathcal{G})\mathbf{x}(t),\;5>t>0,$
with $\mathcal{L}(\mathcal{G})$ denoting the graph Laplacian, given
by: 
\[
\mathcal{L}(\mathcal{G})=\left(\begin{array}{rrrrr}
1 & 0 & 0 & 0 & -1\\
-1 & 1 & 0 & 0 & 0\\
0 & -1 & 1 & 0 & 0\\
0 & 0 & -1 & 1 & 0\\
0 & 0 & 0 & -1 & 1
\end{array}\right).
\]
The edge $\epsilon_{2}$ is removed at time $t_{f}=5$, whence the agents evolution becomes $\dot{\mathbf{x}}(t)=-\mathcal{L}(\bar{\mathcal{G}})\mathbf{x}(t),\; t>5$, where $\mathcal{L}(\bar{\mathcal{G}})$ is the same as $\mathcal{L}({\mathcal{G}})$ above except for its second row which is all zeros. Consequently, the first derivative of $x_{2}(t)$ and the second derivative
of $x_{3}(t)$ exhibit jump discontinuities at $t_{f}=5$, as depicted
in Fig \ref{fig:derivatives}.

For any finite cycle network of arbitrary size, if all derivatives
up to one less than the network size are observed, then any two nodes
offer a solution to not only the detection, but also the isolation
problem. The attention is next shifted to the sensor placement problem
and the performance of the algorithms proposed in Subsections~\ref{sec:coverage}
and \ref{sec:resolution}.

\begin{myexample}\label{example:randomInstances}{Sensor Placement for Detection and Isolation in Simple Random Instances.}\end{myexample}

\begin{figure*}[ht]
\centering 
\begin{subfigure}[b]{0.24\textwidth}
\includegraphics[trim = 10mm 20mm 25mm 8mm,clip,width=50mm]{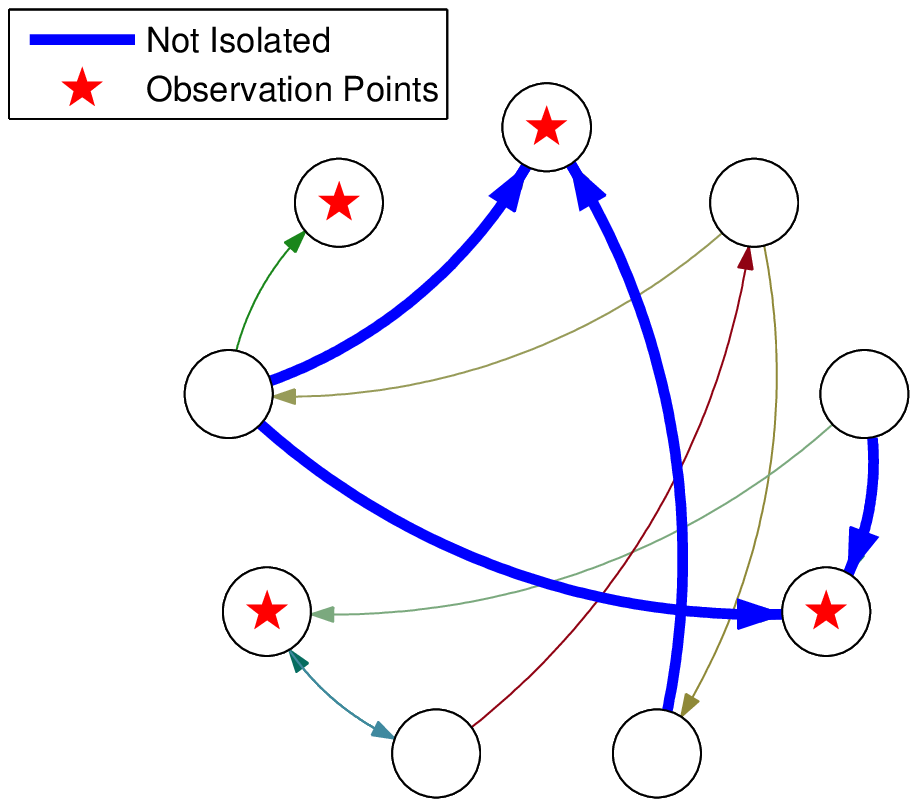} 
\caption{Isolation is impossible} 
\label{fig:DetectionNOTisolation}
\end{subfigure}~
\begin{subfigure}[b]{0.24\textwidth}
\includegraphics[trim = 10mm 20mm 25mm 5mm,clip,width=50mm]{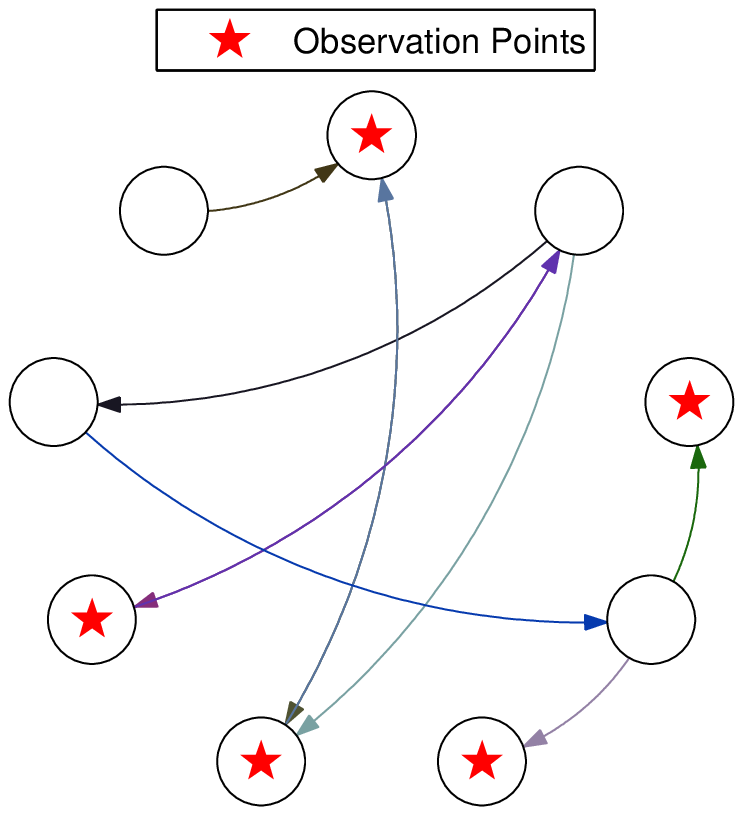} 
\caption{Isolation is possible} 
\label{fig:DetectionANDisolation} 
\end{subfigure}~ 
\begin{subfigure}[b]{0.24\textwidth}
\includegraphics[trim = 10mm 20mm 25mm 2mm,clip,width=50mm]{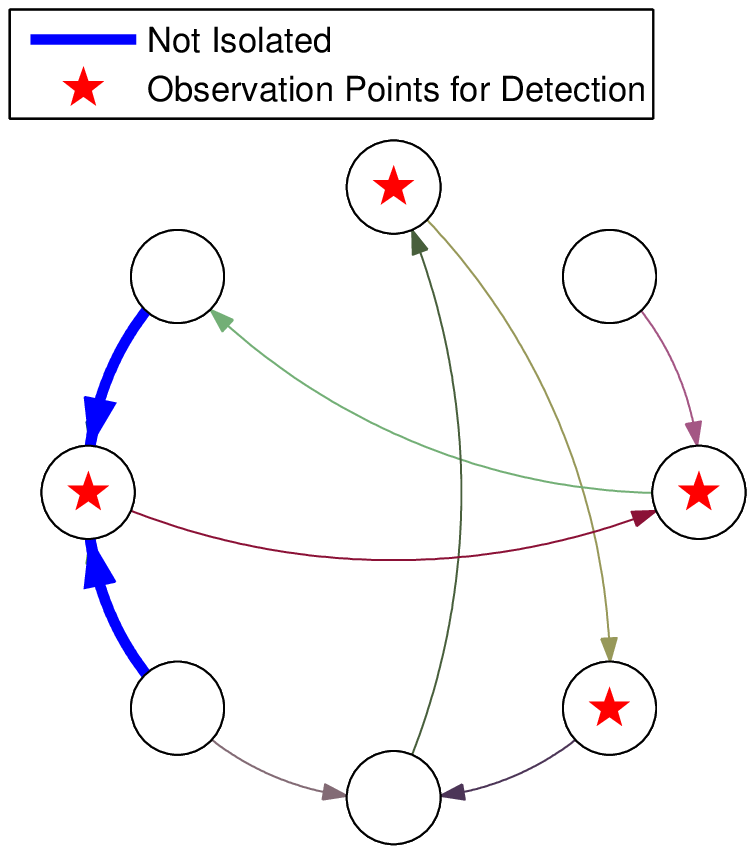}
\caption{Nodes chosen for detection}
\label{fig:detection} 
\end{subfigure}~
\begin{subfigure}[b]{0.24\textwidth}
\includegraphics[trim = 10mm 20mm 25mm 2mm,clip,width=50mm]{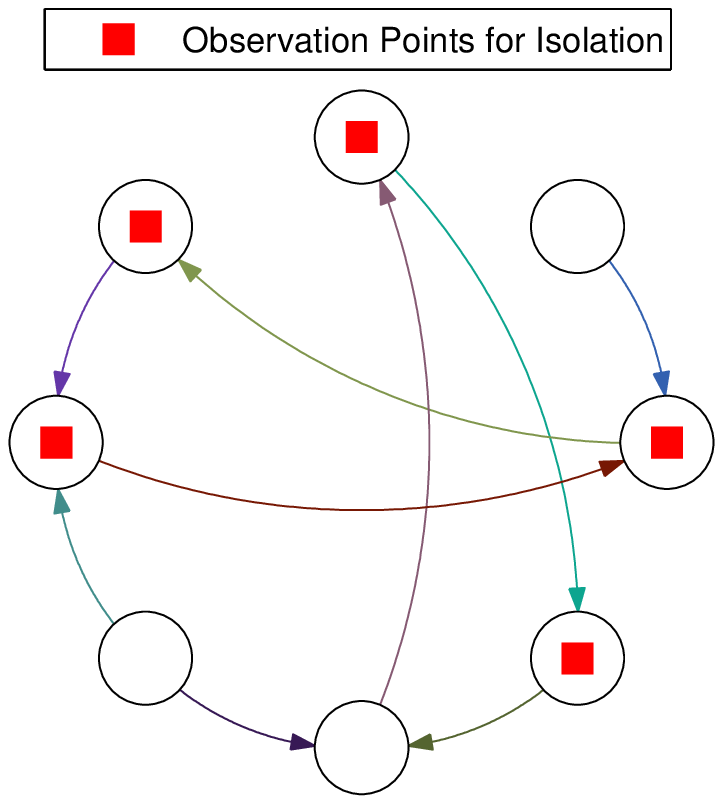} 
\caption{Nodes chosen for isolation}  
\label{fig:isolation}
\end{subfigure}
\caption{ (\subref{fig:DetectionNOTisolation}) Even with all of the nodes selected as observation points, the highlighted links will remain unidentified. (\subref{fig:DetectionANDisolation}) The indicated nodes are enough to achieve both detection and isolation for all of the network links. (\subref{fig:detection}) The highlighted links are not distinguishable using the indicated observation points. (\subref{fig:isolation}) By adding an extra node to the sensor set of (\subref{fig:detection}), all edges become isolated.}
\end{figure*}

Computer codes for Routines \ref{routine:detection} and \ref{routine:isolation}
are developed in MATLAB\textsuperscript{\textregistered{}} and tested
on randomly generated networks. Given the adjacency matrix of the
network, a subroutine computes the matrix $R$ defined at the beginning
of this section, which is then used to compute the outputs of functions
$f_{I}$ and $f_{D}$ for any given subset of nodes, as required for
the implementation of the routines given in Subsections \ref{sec:coverage}
and \ref{sec:resolution}. For an arbitrary graph input to Routines
\ref{routine:detection} and \ref{routine:isolation} several cases
may arise. For the digraph in Fig.~\ref{fig:DetectionNOTisolation},
detection is achieved through the four nodes that are indicated with
stars; however, the highlighted links cannot be isolated using the
output of just these four nodes nodes. In fact, increasing the set
of observation points to include the entire vertex set still does
not reduce the set of indistinguishable edges, and complete isolation
of all links in this digraph is impossible.

The situation in the digraph of Fig.~\ref{fig:DetectionANDisolation}
is reversed, in the sense that with the same set of nodes that are
the output of Routine~\ref{routine:detection} complete isolation
is also achieved. In other words, no extra nodes are needed after
$\mathcal{M}_{I}$ is initialized with $\mathcal{M}_{D}$ in Routine~\ref{routine:isolation}.
For the digraph of Fig.~\ref{fig:detection}, on the other hand,
with the detection output the two highlighted edges cannot be distinguished,
but their status is resolved upon the addition of an extra node in
Fig.~\ref{fig:isolation}.


\begin{myexample}\label{example:experiments}{Computer Experiments with a Random Geometric Graph.}\end{myexample}

\begin{figure}[ht]
\centering \includegraphics[trim = 13mm 0mm 0mm 13mm,clip,width=90mm]{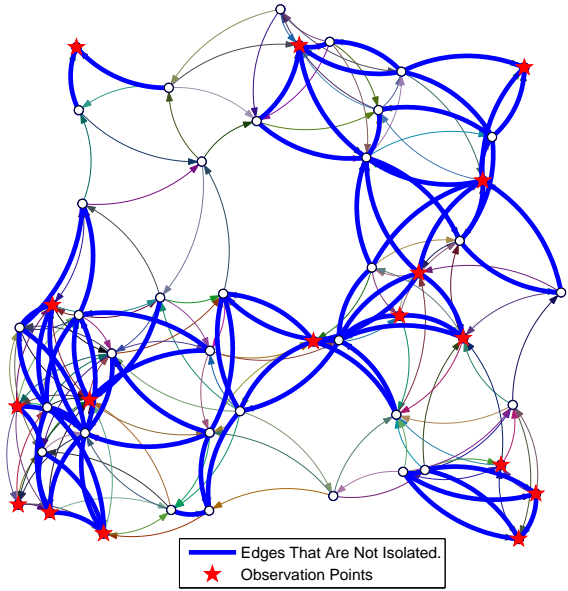}
\caption{By observing upto the $9$-th orders of derivatives in the indicated $17$ nodes, the $75$ edges that are highlighted cannot be isolated, although every edge is detectable; and even with all of the nodes
observed, there still remain $34$ edges that cannot be isolated.}
\label{fig:OrientedRandomGeometricNOTisolated} 
\end{figure}

In the following, the performance of the developed routines
is tested for a random geometric graph model, where the nodes of
the network are randomly and uniformly spread across a bounded region,
and there is an edge wherever a certain distance threshold is met. The orientation of edges in each case is chosen by independent fair coin flips. The graph of Fig.~\ref{fig:OrientedRandomGeometricNOTisolated} 
depicts one such graph instance with $50$ nodes and $200$ unidirectional edges. It follows that a total of $17$ nodes is sufficient for detection, and these $17$ nodes enable the isolation of all but $75$ edges of the digraph, which
are highlighted in Fig.~\ref{fig:OrientedRandomGeometricNOTisolated}. For this directed network, by observing all of the nodes in the network, the cardinality of the set of unresolved links reduces to $34$.

\section{Conclusions}

\label{sec:conc}

In this paper, a method was developed, both analytically and algorithmically,
that enables the designer to detect and isolate
link failures, based on the observed jumps in the derivatives of the
output responses of a subset of nodes in a network of identical LTI
systems. A theorem was presented, which relates the jumps in the derivatives
at the time of failure to the distance of the failed link from the
observation point.  Next a set of efficient algorithms was developed
for sensor placement, which together with the theorem, enables the
designer to determine the existence and location of any link failure
based on the observed jumps. Performance guarantees from the set coverage problem ensure that the proposed algorithms always return a result that is with a $\log|\mathcal{E}| + 1$ factor of the optimum answer, where $E$ is the number of edges in the network.

 { The authors' ongoing research focuses on the development of detection techniques for jumps in high order derivatives using binary hypothesis testing, as well as the investigation of the case where the networked systems are heterogeneous and with bounds on the number of available sensors.}

\appendices

\section{Proofs for the Main Results}\label{proofs}

\subsection{Proof of Theorem~\ref{theo:detection1}: Jump Discontinuities of the Output Derivatives}\label{proof:detectionTheorem}

Given the state space equations in \eqref{eq:StateSPACEFailedSystem}, the transfer matrix from the input $\mathbf{f}_{(i)}\left(t\right)$ to the output $\mathbf{y}_{(p)}\left(t\right)$
can be written as: 
\begin{align}
& \left[\mathbb{H}(s)\right]_{(p)(i)}  =\left(\mathbf{e}_{p,N}^{T}\otimes C\right) \times\ldots \nonumber \\ & \left[sI_{Nd}-(I_{N}\otimes A+G\otimes B\Gamma C)\right]^{-1}\left(\mathbf{e}_{i,N}\otimes B\right)\\
 & =\left(\mathbf{e}_{p,N}^{T}\otimes C\right) \times\ldots \nonumber \\ & \left[\left(I_{N}\otimes\left(sI_{d}-A\right)\right)-\left(G\otimes B\Gamma C\right)\right]^{-1}\left(\mathbf{e}_{i,N}\otimes B\right)\label{eq:transferMatrix}
\end{align}
Pre- and post-multiplication by $\left[I_{N}\otimes\left(sI_{d}-A\right)^{-1}\right]$
yields: 
\begin{align}
& \left[\mathbb{H}(s)\right]_{(p)(i)}=\left(\mathbf{e}_{p,N}^{T}\otimes C\right)\left[I_{N}\otimes\left(sI_{d}-A\right)^{-1}\right] \times\ldots \nonumber \\ & \left\{ I_{Nd}-\left[G\otimes B\Gamma C\left(sI_{d}-A\right)^{-1}\right]\right\} ^{-1}\left(\mathbf{e}_{i,N}\otimes B\right)\label{eq:eq:transferMatrix2}
\end{align}
The matrix inverse can be expanded as follows: 
\begin{align}
&\left[\mathbb{H}(s)\right]_{(p)(i)}=\left(\mathbf{e}_{p,N}^{T}\otimes C\right)\left[I_{N}\otimes\left(sI_{d}-A\right)^{-1}\right] \times\ldots \nonumber \\ &\left\{ I_{Nd}+\sum_{l=1}^{\infty}{\left[G\otimes B\Gamma C\left(sI_{d}-A\right)^{-1}\right]}^{l}\right\}  \left(\mathbf{e}_{i,N}\otimes B\right) \nonumber
\end{align}
Write
\begin{align}
{\left[G\otimes B\Gamma C\left(sI_{d}-A\right)^{-1}\right]}^{l}={G^{l}\otimes{\left(B\Gamma C\left(sI_{d}-A\right)^{-1}\right)}^{l}}, \nonumber
\end{align} and apply Lemma~\ref{lem:numWalksLemma} for the $pi-$th block of the center matrix,  to get: 
\begin{align}
\left[\mathbb{H}(s)\right]_{(p)(i)}  =  \sum_{l=\textrm{dist}(\nu_{i},\nu_{p})}^{\infty} \{ \left[G^{l}\right]_{pi}C\left(sI_{d}-A\right)^{-1} \times \ldots & \nonumber \\  \left(B\Gamma C\left(sI_{d}-A\right)^{-1}\right)^{l}B\}&.
\end{align}

Using $H(s)=C(s{I_d}-A)^{-1}B$, the above can be rewritten as: 
\begin{align}
\left[\mathbb{H}(s)\right]_{(p)(i)}\Gamma=\sum_{l=\textrm{dist}(\nu_{i},\nu_{p})}^{\infty}\left[G^{l}\right]_{pi}\left[H(s)\Gamma\right]^{l+1}.\label{eq:TransferMatrixPIblock}
\end{align}

From \eqref{outputEquations} the Laplace transform of the output at node $p$ is given by: 
\begin{align}
&\mathbf{\hat{y}}_{(p)}(s) =  \left(\mathbf{e}_{p,N}^{T}\otimes I_{o}\right)\mathbb{H}(s) \left(\mathbf{\hat{f}}(s)+\mathbf{W}(s)\right) \label{eq:outputResponseLaplaceTransform} \\ 
& +  {\left(\mathbf{e}_{p,N}^{T}\otimes C\right)\left[sI_{Nd}-(I_{N}\otimes A+G\otimes B\Gamma C)\right]^{-1}}\mathbf{x}\left(t_{0}\right)e^{-t_{0}s} \nonumber
\end{align} { Note that Laplace transforms of outputs in the the healthy and faulty systems differ only in the term $\mathbf{\hat{f}}(s)$ appearing in \eqref{eq:outputResponseLaplaceTransform}.} To measure the jump in the $k$-th derivative of $\mathbf{{y}}_{(p)}\left(t\right)$ due to the link failure stimulated by the virtual input $\mathbf{f}\left(t\right)$, we apply the initial value theorem to the $k$-th derivatives of the time shifted outputs for the healthy and faulty systems and calculate their difference to get:
\begin{align}
\boldsymbol{\Delta}(p,k) & = \mbox{lim}_{s\to\infty}s^{k+1}e^{t_{f}s}\left(\mathbf{e}_{p,N}^{T}\otimes I_{o}\right)\mathbb{H}\left(s\right)\mathbf{\hat{f}}\left(s\right) \nonumber \\
 & =\mbox{lim}_{s\to\infty}s^{k+1}e^{t_{f}s}\left[\mathbb{H}(s)\right]_{(p)(i)}\mathbf{\hat{f}}_{(i)}(s), \nonumber
\end{align} { which can be decomposed as
\begin{align}
&\boldsymbol{\Delta}(p,k) = \label{eq:derivativeJumps} \\ & \left(\mbox{lim}_{s\to\infty}s^{k}\left[\mathbb{H}(s)\right]_{(p)(i)}\right)\left( \mbox{lim}_{s\to\infty}se^{t_{f}s}\mathbf{\hat{f}}_{(i)}(s) \right). \nonumber
\end{align}
The remainder of the proof is in calculating the two limits appearing in \eqref{eq:derivativeJumps}. We use Laplace domain techniques along with facts from complex analysis to calculate $\mbox{lim}_{s\to\infty}se^{t_{f}s}\mathbf{\hat{f}}_{(i)}(s)$, which we shall see is contributing a constant multiplier factor to the quantity $\boldsymbol{\Delta}(p,k)$. On the other hand, the calculation of $\mbox{lim}_{s\to\infty}s^{k}\left[\mathbb{H}(s)\right]_{(p)(i)}$ is based primarily on Lemma~\ref{lem:numWalksLemma} and that is where the inter-nodal distance relations come into play. 

We begin by taking the Laplace transform of $\mathbf{f}_{(i)}\left(t\right)$ in
\eqref{eq:ReplicantInput}: 
\begin{align}
\mathbf{\hat{f}}_{(i)}(s)=  -g_{ij}\Gamma C \int_{0}^{+\infty}\mathbf{x}_{(j)}(t)\bar{H}_{t_f}(t)e^{-st}\, d t,  \label{eq:LaplaceTransformOfFaultReplicantInput}
\end{align} where $\bar{H}_{t_f}(t)$ is the shifted Heaviside step function given by $\bar{H}_{t_f}(t) = 0$ for $t< t_f$ and $\bar{H}_{t_f}(t) = 1$ otherwise. The Laplace transform of $\bar{H}_{t_f}(\mathord{\cdot})$ is given by $\hat{H}_{t_f}(s) = e^{-t_f s}/{s}$ for $s \in \mathbb{C}$ having positive real part. Using the contour integral for inverse Laplace transform \cite[Chapter 7]{brown2009complex}, we can write 
\begin{align}
\bar{H}_{t_f}(t) = \frac{1}{2\pi J}\int_{\sigma-j\infty}^{\sigma+j\infty}\frac{e^{-t_f \zeta}}{{\zeta}}\, d\zeta, \label{eq:LaplaceInverse}
\end{align} for $\sigma$ large enough such that the vertical line from ${\sigma-j\infty}$ to ${\sigma+j\infty}$ lies to the right of all singularities of the integrand ${e^{-t_f \zeta}}/{{\zeta}}$. Next replacing \eqref{eq:LaplaceInverse} in $\mathbf{L}(s):=\int_{0}^{+\infty}\mathbf{x}(t)\bar{H}_{t_f}(t)e^{-st}\, d t$ and switching the order of the integrals yields 
\begin{align}
\mathbf{L}(s) =  \frac{1}{2\pi J} \int\limits_{\sigma-J\infty}^{\sigma+J\infty}\frac{e^{-t_f \zeta}}{{\zeta}}\int\limits_{0}^{+\infty} \mathbf{x}_{(j)}(t)e^{-(s-\zeta)t}\, dt\, d\zeta.  \label{eq:LaplaceOrderExchanged}
\end{align} Using the Laplace transform 
\begin{align}
\mathbf{\hat{x}}_{(j)}(s-\zeta) = \int_{0}^{\infty} \mathbf{x}_{(j)}(t)e^{-(s-\zeta)t}\, dt,
\end{align} and the change of variable $\eta = s - \zeta$, \eqref{eq:LaplaceOrderExchanged} can be rewritten as 
\begin{align}
\mathbf{L}(s) & = \frac{1}{2\pi J}\int\limits_{\sigma'-J\infty}^{\sigma'+J\infty} \frac{e^{-t_f (s-\eta)}} {\eta-s} \mathbf{\hat{x}}_{(j)}(\eta) \, d\eta \nonumber \\  & =   \frac{e^{-t_f s}}{2\pi J}\int\limits_{\sigma'-J\infty}^{\sigma'+J\infty}\frac{e^{t_f \eta}}{{\eta -s}}\mathbf{\hat{x}}_{(j)}(\eta)\, d\eta, 
\label{eq:VariablesShifted}
\end{align} for $\sigma'$ large enough. Wherefore we get that
\begin{align}
& \lim_{s\to\infty}se^{t_fs}\mathbf{L}(s) = \frac{1}{2\pi J}\int\limits_{\sigma'-J\infty}^{\sigma'+J\infty}\lim_{s\to\infty}\frac{se^{t_f \eta}}{{\eta-s}}\mathbf{\hat{x}}_{(j)}(\eta)\, d\eta \nonumber \\ &= \frac{-1}{2\pi J} \int\limits_{\sigma'-J\infty}^{\sigma'+J\infty}e^{t_f \eta} \mathbf{\hat{x}}_{(j)}(\eta)\, d\eta  = -\mathbf{{x}}_{(j)}(t_f), \label{eq:Laplace}
\end{align} exchanging the limit and integral in the first equality justifiable through the dominated convergence, and the last equality following by the inverse Laplace transform. Using \eqref{eq:Laplace} in \eqref{eq:LaplaceTransformOfFaultReplicantInput} thus establishes that 
\begin{align}
\mbox{lim}_{s\to\infty}se^{t_{f}s}\mathbf{\hat{f}}_{(i)}(s) = g_{ij}\Gamma C\mathbf{{x}}_{(j)}(t_f). \label{eq:FirstLimit}
\end{align}
}
Next for the limit term, $\lim_{s\to\infty}s^{k}\left[\mathbb{H}(s)\right]_{(p)(i)}$, appearing in \eqref{eq:derivativeJumps} we have
\begin{align}
&\lim_{s\to\infty}s^{k}\left[\mathbb{H}(s)\right]_{(p)(i)}  \nonumber \\ & = \mbox{lim}_{s\to\infty}s^{k}g_{ij}\left[\sum_{l=\textrm{dist}(\nu_{i},\nu_{p})}^{\infty}\left[G^{l}\right]_{pi}\left[H\left(s\right)\Gamma\right]^{l+1}\right] \nonumber \\ & = g_{ij}\left[\sum_{l=\textrm{dist}(\nu_{i},\nu_{p})}^{\infty}\left[G^{l}\right]_{pi}\mbox{lim}_{s\to\infty}s^{k}\left[H\left(s\right)\Gamma\right]^{l+1}\right].\label{eq:dominatedConvergence2}
\end{align} However, with $k=r(\textrm{dist}(\nu_{i},\nu_{p})+1)$ in \eqref{eq:dominatedConvergence2}, it follows by the choice of $r$ being the relative degree that $\mbox{lim}_{s\to\infty}s^{k}\left[H(s)\Gamma\right]^{l+1}:=Q\in R^{o\times o}$
for $l=\textrm{dist}(\nu_{i},\nu_{p})$ and $\mbox{lim}_{s\to\infty}s^{k}\left[H(s)\Gamma\right]^{l+1} = Z_{o}$
for $l>\textrm{dist}(\nu_{i},\nu_{p})$; while for $k<r(\textrm{dist}(\nu_{i},\nu_{p})+1)$,
$\mbox{lim}_{s\to\infty}s^{k}\left[H(s)\Gamma\right]^{l+1}=Z_{o}$ for
all $l\geqslant\textrm{dist}(\nu_{i},\nu_{p})$. In particular, we have that
\begin{align}
&\lim_{s\to\infty}s^{k}\left[\mathbb{H}(s)\right]_{(p)(i)} \label{eq:secondLimit} \\  &= \left\{
\begin{array}{ll}
\left[G^{k}\right]_{pi} Q,  & \mbox{if } k = r(\textrm{dist}(\nu_{i},\nu_{p})+1), \\
Z_{o}, & \mbox{if } k<r(\textrm{dist}(\nu_{i},\nu_{p})+1),
\end{array}
\right. \nonumber
\end{align} and the claim follows upon replacing \eqref{eq:FirstLimit} and \eqref{eq:secondLimit} in \eqref{eq:derivativeJumps}. \hfill{}{\scriptsize $\blacksquare$}{\scriptsize \par}

\subsection{Claim~\ref{claim:f_D}: Submodularity of Coverage Function}
For the proof we need the additional concept of a coverage function, and we use a known result form the theory of submodular set functions that coverage functions are submodular \cite{CombOpt}. Given a collection of subsets of edges $E_{1},E_{2},\ldots,E_{N}\subset\mathcal{E}$, where each $E_q$ is associated with a node $\nu_q$, the coverage function $f_{C}:\mathcal{P}\left(\mathcal{V}\right)\to\mathbb{R}_{+}$ is defined as $f_{C}\left(S\right)=\left|\cup_{\nu_{q}\in S}E_{q}\right|$ for any $S\subseteq\mathcal{V}$. Given a node $\nu_{q}$, let $C(\nu_{q})=\left\{ \epsilon\in\mathcal{E}:\left(q,\epsilon\right)\notin\mathcal{R}_{0}\right\}$ be the correspondence that for each node $\nu_q$ in the network gives the set of edges whose failure does induce a jump in any of the first $z$ derivatives of $\mathbf{y}_{q}\left(t\right)$.  We have that $f_{D}(\mathcal{M})=\left|\cap_{\nu_{q}\in\mathcal{M}}C_{q}^{c}\right|=\left|\left(\cup_{\nu_{q}\in\mathcal{M}}C_{q}\right)^{c}\right|=\left|\mathcal{E}\right|-\left|\left(\cup_{\nu_{q}\in\mathcal{M}}C_{q}\right)\right|$, where $^{c}$ denotes the set complement w.r.t. $\mathcal{E}$. The claim now follows upon noting that the latter term is a coverage function. \hfill{}{\scriptsize $\blacksquare$}{\scriptsize \par}

\subsection{Claim~\ref{claim:f_I}: Submodularity of Resolution Function}
Consider any two subsets of nodes $\bar{\mathcal{M}}$ and $\hat{\mathcal{M}}$ such that  $\bar{\mathcal{M}}\subset\hat{\mathcal{M}}\subset\mathcal{V}$ and a vertex $\nu_q \in \mathcal{V}\fgebackslash\left(\bar{\mathcal{M}}\cup\hat{\mathcal{M}}\right)$. Note that $\bar{\mathcal{M}}\subset\hat{\mathcal{M}}$ implies that $\mathcal{U}(\hat{\mathcal{M}}) \subset \mathcal{U}(\bar{\mathcal{M}})$ so that $f_{I}(\hat{\mathcal{M}}) \leqslant f_{I}(\bar{\mathcal{M}})$. In particular, $ f_{I}(\bar{\mathcal{M}})  \geqslant f_{I}(\bar{\mathcal{M}}\cup\{\nu_q\})$. Now, take any $\epsilon \in \mathcal{U}(\hat{\mathcal{M}})$ and note that there exists an $\hat{\epsilon} \in \mathcal{U}(\hat{\mathcal{M}})$ such that $\mathcal{I}(\mathcal{M},\epsilon) = \mathcal{I}(\mathcal{M},\hat{\epsilon})$. Now if all such $\epsilon$ and $\hat{\epsilon}$ are in the same relations $\mathcal{R}_k, k \in [z]\cup\{0\}$ with the vertex $\nu_q$, then $f_{I}(\hat{\mathcal{M}}\cup\{\nu_q\})- f_{I}(\hat{\mathcal{M}}) = 0  \geqslant f_{I}(\bar{\mathcal{M}}\cup\{\nu_q\})- f_{I}(\bar{\mathcal{M}})$. Next consider the case where $\mathcal{U}(\hat{\mathcal{M}})\fgebackslash\mathcal{U}(\hat{\mathcal{M}}\cup\{\nu_q\}) \neq \varnothing$. Any edge $\epsilon \in \mathcal{U}(\hat{\mathcal{M}})\fgebackslash\mathcal{U}(\hat{\mathcal{M}}\cup\{\nu_q\})$ is such that for all edges $\hat{\epsilon} \in \mathcal{U}(\hat{\mathcal{M}})$ having $\mathcal{I}(\mathcal{M},\epsilon)=\mathcal{I}(\mathcal{M},\hat{\epsilon})$, $\epsilon$ and $\hat{\epsilon}$ satisfy different binary relations with the added vertex $\nu_q$. But then since both $\epsilon$, $\hat{\epsilon} \in \mathcal{U}(\bar{\mathcal{M}})$ and are in different relations with $\nu_q$ it follows that $\epsilon$, $\hat{\epsilon} \in \mathcal{U}(\bar{\mathcal{M}})\fgebackslash\mathcal{U}(\bar{\mathcal{M}}\cup\{\nu_q\})$. Hence, we have shown that $\mathcal{U}(\hat{\mathcal{M}})$ $\fgebackslash$ $\mathcal{U}(\hat{\mathcal{M}}\cup\{\nu_q\})$ $\subset$ $\mathcal{U}(\bar{\mathcal{M}})$ $\fgebackslash$ $\mathcal{U}(\bar{\mathcal{M}}\cup\{\nu_q\})$ so that again $|\mathcal{U}(\hat{\mathcal{M}})$ $\fgebackslash$ $\mathcal{U}(\hat{\mathcal{M}}\cup\{\nu_q\})| = f_{I}(\hat{\mathcal{M}}) - f_{I}(\hat{\mathcal{M}}\cup\{\nu_q\}) \leqslant f_{I}(\bar{\mathcal{M}}) - f_{I}(\bar{\mathcal{M}}\cup\{\nu_q\}) =  |\mathcal{U}(\bar{\mathcal{M}})$ $\fgebackslash$ $\mathcal{U}(\bar{\mathcal{M}}\cup\{\nu_q\})|$, and the property of submodularity for $-f_I(\cdot)$ follows by definition.  \hfill{}{\scriptsize $\blacksquare$}{\scriptsize \par}

\bibliographystyle{IEEEtran}
\bibliography{refDistinguishability,newRef}

\begin{biography}
[{%
\includegraphics[width=1in,height=1.25in,clip,keepaspectratio]{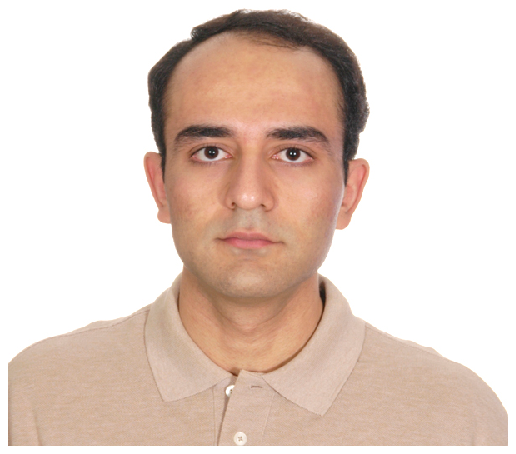}
}] {Mohammad Amin Rahimian} is a recipient of gold medal in 2004 Iran National Chemistry Olympiad. He was awarded an honorary admission to Sharif University of Technology, where he received his B.Sc. in Electrical Engineering-Control. In 2012, he received his M.A.Sc. in Electrical and Computer Engineering at Concordia University, and he is currently a PhD student at the GRASP Laboratory, University of Pennsylvania. His research interests include systems dynamics and network theory, control of multi-agent networks, localization of mobile sensors, and tuning of fractional controllers.
\end{biography}

\begin{biography}
[{%
\includegraphics[width=1in,height=1.25in,clip,keepaspectratio]{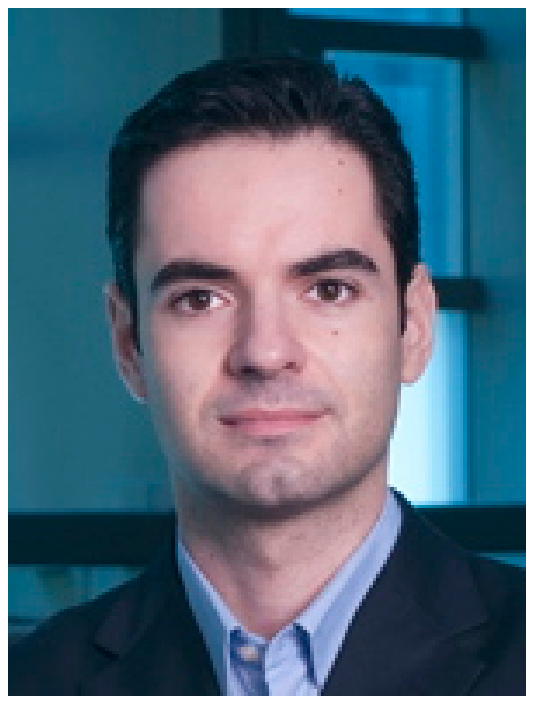}
}] {Victor M. Preciado} received the Ph.D. degree in electrical engineering
and computer science from the Massachusetts Institute of Technology,
Cambridge, in 2008.

He is currently the Raj and Neera Singh Assistant Professor of Electrical
and Systems Engineering at the University of Pennsylvania. He is a member of the Networked and Social Systems Engineering (NETS) program and the Warren Center for Network and Data Sciences. His research interests include network science, dynamic systems, control theory, complexity, and convex optimization with applications in social networks, technological infrastructure, and biological systems.
\end{biography}

\end{document}